\documentclass{amsproc}
\usepackage{amssymb}

\newtheorem {theorem}{Theorem}[section]
\newtheorem {lemma}[theorem]{Lemma}
\theoremstyle{definition}

\theoremstyle{remark}

\begin{document}

\newcommand{\be}{\begin{equation}}
\newcommand{\ba}{\begin{eqnarray}}
\newcommand{\ea}{\end{eqnarray}}
\newcommand{\n}{\noindent}
\newcommand{\f}{\frac}

\title{ Schr\"{o}dinger Equation with  Moving Point Interactions in
Three Dimensions}

\author{ G.F. Dell'Antonio}
\address{Dipartimento di Matematica, Universita' di Roma "La 
Sapienza",
Italy and  Laboratorio Interdisciplinare, S.I.S.S.A., Trieste, Italy.}
\email{dellantonio@mat.uniroma1.it}

\author{ R. Figari}
\address{Dipartimento di Scienze Fisiche, Universita' di Napoli, 
Italy.}
\email{figari@na.infn.it}

\author{ A. Teta}
\address{Dipartimento di Matematica, Universita' di Roma "La 
Sapienza",
Italy.}
\email{teta@mat.uniroma1.it}

\vspace{1.5cm}
\begin{abstract}
We consider the motion of a non relativistic quantum particle in
$R^{3}$ subject to $n$ point interactions which are moving on
given smooth trajectories. Due to the singular character of the
time-dependent
interaction, the corresponding Schr\"{o}dinger equation does not  have
solutions in a strong sense and, moreover,
standard perturbation techniques cannot be used. Here we prove that, 
for
smooth
initial data,  there is a
unique weak solution by reducing the problem to the solution of
a Volterra integral equation involving only the time variable.
It is also shown  that the
evolution operator   uniquely extends to a   unitary operator in
$L^{2}(R^{3})$.
\end{abstract}

\maketitle

\hfill {\em Dedicated to Sergio Albeverio}

\vspace{1cm}

\section{Introduction}

\vspace{0.5cm}

We consider the Schr\"{o}dinger equation in $R^{3}$ with an 
interaction
supported
by $n$ points which are moving on  preassigned smooth paths.
More precisely let $\alpha  = (\alpha_{1}, \ldots ,
\alpha_{n})$ be a vector in $R^{n}$
  and  let $y(t) =(y_{1}(t), \ldots ,y_{n}(t))$ be $n$
  given smooth non intersecting trajectories  in $R^{3}$. For  $t \in
R$, let   $H_{\alpha,y(t)}$ be the Schr\"{o}dinger operator in
$L^{2}(R^{3})$ with point interactions supported at
$y(t)$ and with strength $\alpha$. We recall below the explicit 
definition
of $H_{\alpha,y(t)}$.

\n
We are interested in the   non-autonomous evolution problem

\be
i \;  \f{\partial \psi_{s} (t)}{\partial t} = H_{\alpha,y(t)}
\psi_{s}(t), \;\;\;\; \psi_{s}(s) = f
\end{equation}

\n
where $s$ is an  arbitrary initial time and $f$ is some (possibly
smooth) initial datum.

\n
An existence theorem for the solution of problem (1) cannot be given 
using
the
standard theory of
non-autonomous evolution problems (see e.g. \cite{Ya}) because of the 
strong
dependence
on time of the operator domain, and in fact even of the form domain
of $H_{\alpha,y(t)}$. 
Note that the case of point interactions at fixed positions with
time-dependent strengths (see \cite{SY},\cite{Y}) is less singular 
since 
the form domain is constant.

\n
As we shall see, problem (1) does not have solutions in
a strong sense. The  reason is that, even for very smooth initial
datum, the solution exhibits an additional singularity at the position
of the moving  points and then it does not belong to the operator
domain.

\n
In a previous paper (\cite{DFT2}) we studied the corresponding 
problem for
the heat equation. For each $u_{0} \in D(H_{\alpha, y(0)})$ we proved
existence and uniqueness of a strong solution, i.e. of a function
$u(t)$ belonging for each $t>0$ to  $D(H_{\alpha,y(t)})$,
satisfying in the  $L^{2}$-sense  the equation

\be
\f{\partial u(t)}{\partial t} = - H_{\alpha,y(t)} u(t) , \;\;\;\;
u(0)=u_{0}
\end{equation}

\n
The proof exploited the smoothing properties of the heat kernel  and 
it cannot be generalized to the Schr\"{o}dinger case.

\n
In this paper we show  that, when interpreted in a suitable weak
sense, problem (1) has a unique solution.

\n
More precisely, let ${\mathcal B}_{y(t)} (\cdot , \cdot )$ be the 
bilinear
form associated to $H_{\alpha, y(t)}$ and let $V_{t}$ be its domain
(which depends on $y(t)$). Let

\be
C_{y(t)}^{\infty} \equiv  C_{0}^{\infty}(R^{3} \setminus
\{y(t)\} )
\end{equation}

\n
and notice that $C_{y(t)}^{\infty} \subset D(H_{\alpha,y(t)})$ (see 
(5)).

\n
We shall prove that for all $f \in C_{y(s)}^{\infty}$ there is a
unique solution of the equation

\be
i \left( v(t), \f{\partial \psi_{s}(t)}{\partial t} \right)  =
{\mathcal B}_{y(t)}(v(t) \psi_{s}(t) ), \;\;\; \psi_{s}(s) =f
\end{equation}

\n
for all $v(t) \in V_{t}$. Moreover $\psi_{s}(t)$ has a natural
representation (see (14)).

\n
The maps $f \rightarrow \psi_{s}(t)$, $s,t \in R$, are isometries  and
extend by continuity to unitary maps $U(t,s)$ in $L^{2}(R^{3})$. The
maps $U(t,s)$ are continuous in $s,t$ in the strong operator topology,
and therefore define a time-dependent dynamical system in
$L^{2}(R^{3})$, with generator $H_{\alpha, y(t)}$ at time $t$.

\n
Notice that due to the assumptions on the initial data we do not
define a flow on $V_{t}$. We conjecture however that indeed problem
(4)
defines a flow in $V_{t}$, continuous  with respect to the Banach
topology defined on $V_{t}$ by the bilinear form ${\mathcal 
B}_{y(t)}$.

\n
We consider the solution of problem (4) as the first  step  in the 
study
of
the motion
of a quantum particle (e.g. a neutron) in a fluid, regarded as
 an assembly of $n$  classical   particles, each of which acts
through a potential of very short range  and therefore can be
considered as a
point
interaction. The limit $n$ going to infinity for the 
 case of the heat equation  was studied
by us in \cite{DFT1}.

\n
The results presented here  are also a preliminary step in the
analysis of a class of  nonlinear models in which  the motion of the 
$n$
classical
 particles is not preassigned but rather determined by the
interaction with the quantum particle.

\vspace{1cm}

\section{Definitions, motivations and statement of the results}

\vspace{0.5cm}

We have denoted by $H_{\alpha, y(t)}$ the Schr\"{o}dinger operator in
$L^{2}(R^{3})$ with point interactions of strength $\alpha
=(\alpha_{1}, \ldots , \alpha_{n})$ placed on the points with
coordinates $y(t)=(y_{1}(t), \ldots , y_{n}(t))$. For the sake of
simplicity we have assumed that the strengths are constant, and we 
shall
omit them in the labels from now on. The extension to the case where
also the strength of the interactions depends on time is
straightforward, since this dependence on time does not alter the
form domain.

\n
The operator $H_{y(t)}$ is self-adjoint, bounded below, with domain
and action given respectively for each value of $t$ by

\ba
& &D(H_{y(t)}) = \left\{ u(t) \in L^{2}(R^{3}) \;|\; u(t) =
\phi (t) +  \sum_{k=1}^{n} q_{k}(t) G(\cdot - y_{k}(t)),\right.
\nonumber\\
& & \phi(t) \in H^{2}_{loc}(R^{3}), \; \Delta \phi(t) \in
L^{2}(R^{3}), \; q_{1}(t), \ldots , q_{n}(t) \in C, \nonumber\\
& & \left. \lim_{|x-y_{k}(t)| \rightarrow 0}
\left[ u(x,t) - q_{k}(t) G(x-y_{k}(t) \right] = \alpha_{k} q_{k}(t), 
\;
k=1, \ldots n \right\}
\ea

\be
H_{\alpha,y(t)}  u(t) = - \Delta
\phi(t)
\end{equation}

\n
Here $H^{m}(R^{3})$ is the standard  Sobolev space,
$C$ denotes the set of complex numbers
and $G$ is the Green's function

\be
G(x-x') = (- \Delta )^{-1}(x-x') =
\frac{1}{4 \pi |x-x'|}, \;
\end{equation}

\n
It is clear from (5) that the operator domain consists of functions 
with
a regular part $\phi(t)$ plus the "potential" produced by the  "point
charges" $q_{k}(t)$. The limit in (5) is regarded as a boundary 
condition
satisfied by $u(t)$ at $y(t)$.

\n
We refer to \cite{AGH-KH} for a complete analysis of this kind of
hamiltonians.

\n
Denote by ${\mathcal F}_{ y(t)}$,$D({\mathcal F}_{y(t)})$ the
 closed and bounded below   quadratic form associated to
$H_{y(t)}$ and
let ${\mathcal B}_{y(t)}$ the corresponding bilinear form.

\n
One has (see
\cite{T} for 
details)

\ba
& &D({\mathcal F}_{y(t)}) = \{ u(t) \in L^{2}(R^{3}) \; | \;  u(t) =
\phi(t) + \sum_{k=1}^{n} q_{k}(t) G(\cdot - y_{k}(t)),\nonumber\\
& & \phi(t) \in H^{1}_{loc}(R^{3}), \; |\nabla \phi (t)| \in 
L^{2}(R^{3})
\}\\
& &{\mathcal F}_{y(t)}(u(t)) = \int_{R^{3}} dx | \nabla \phi(x,t) 
|^{2}
  + \sum_{k=1}^{n} \alpha_{k} |q_{k}(t)|^{2} \nonumber\\
& &- \sum_{k,j=1,k \neq
  j}^{n} G(y_{k}(t) - y_{j}(t) ) \bar{q}_{k}(t) q_{j}(t)
\ea

\n
where   $\bar{z}$ denotes the complex conjugate of $z \in C$.   

\n
Notice that $|\nabla G| \not\in  L^{2}(R^{3})$, and therefore the
decomposition in (8) is unique. We also emphasize that $D( {\mathcal
F}_{y(t)})$ is strictly larger than the form domain of the laplacian.
\n
To simplify the notation we denote by $V_{t}$ the Hilbert space
$D({\mathcal F}_{y(t)})$ equipped with the scalar product

\be
<v(t),u(t)> =   {\mathcal B}_{y(t)}(v(t),u(t)) + \beta (v(t),u(t))
\end{equation}

\n
 where $\beta > - \inf \sigma(H_{y(t)})$.

\n
We also introduce the dual space $V^{*}_{t}$ of $V_t$ with respect to 
the
$L^{2}$-scalar product and denote by $(\xi (t), \eta(t) )$ the
corresponding duality, $\xi(t) \in V_t$, $\eta(t) \in V^{*}_{t}$.

\n
Finally we define the following set of smooth curves in $R^{3}$

\ba
& &{\mathcal M} = \{ y \equiv (y_{1}, \ldots , y_{n}) \;|\; y_{j} 
\;:\; R
\rightarrow R^{3} , \; y_{j} \; \mbox{is of class}\; C^{3},  
\nonumber\\
& & j=1, \ldots n, \; \inf_{j
\neq l} \inf_{t \in R} |y_{j}(t) - y_{l}(t)| \geq a>0 \}
\ea

\n
With these notation our main results are the following.

\vspace{0.5cm}
\begin{theorem}
\label{Th 1} Let $y \in {\mathcal M}$, $s \in R$  and
$f \in C_{y(s)}^{\infty}$. Then there exists a
unique $\psi_{s}(t) \in V_t$, $t \in R$, such that $\f{\partial 
\psi_{s} (t
)}{\partial t} \in V^{*}_{t}$ and

\ba
& &i \left( v(t), \f{\partial \psi_{s} (t)}{\partial t} \right) \;  =
{\mathcal B}_{y(t)}(v(t),\psi_{s} (t) ) \;\;\;\;\; \forall v(t) \in 
V_t \\
& &\psi_{s}(s)= f
\ea

\n
Moreover $\psi_{s}(t)$ has the following representation for $t>s$

\be
\psi_{s}(t) = U_{0}(t-s) f + i \sum_{j=1}^{n} \int_{s}^{t} d\tau \; 
U_{0}(t-
\tau; \cdot -
y_{j}(\tau)) q_{j}(\tau)
\end{equation}

\n
where $U_{0}(t)$ is the free unitary group defined by the  kernel

\be
U_{0}(t; x-x') = e^{i \Delta t}(x-x') =
\f{e^{i \f{|x-x'|^{2}}{4t}}}{(4 \pi it)^{3/2}}
\end{equation}

\n
and the charges $q_{j}(t)$ satisfy the Volterra integral equation

\ba
& &q_{j}(t) + \alpha_{j} \f{4 \sqrt{\pi}}{\sqrt{-i}}
\int_{s}^{t} d\tau \f{q_{j}(\tau)}{\sqrt{t- \tau}} + \int_{s}^{t} 
d\tau
q_{j}(\tau)
C_{j}(t,\tau) \nonumber\\
& &+ \sum_{l=1, l\neq j}^{n} \int_{s}^{t} d \tau q_{l}(\tau)
D_{jl}(t, \tau )    = \f{4 \sqrt{\pi}}{\sqrt{-i}}
\int_{s}^{t} d\tau \f{(U_{0}(\tau - s)
f)(y_{j}(\tau))}{\sqrt{t-\tau}}\nonumber\\
& &
\ea

\n
where

\ba
& &C_{j}(t,\tau)= - \f{1}{\pi} \int_{\tau}^{t} d\sigma \f{1}{\sqrt{t -
\sigma}\sqrt{\sigma - \tau}}
\left(i A_{jj}(\sigma, \tau) + \f{dB_{jj}}{d \tau} (\sigma,\tau) 
\right.
\nonumber\\
& &+ \left. \f{B_{jj}(\sigma,\tau)  - 1}{2
(\sigma  - \tau)} \right)
\ea

\be
A_{jl}(t,\tau)=  \f{(y_{j}(t) -y_{l}(\tau)) \cdot
\dot{y}_{l}(\tau)}{2(t-
\tau)} \f{1}{w^{3}_{jl}
(t,\tau)} \int_{0}^{w_{jl}(t, \tau)} dz  \; z^{2} e^{i z^{2}}
\end{equation}

\be
B_{jl}(t,\tau)=  \f{1}{w_{jl}(t, \tau)} \int_{0}^{w_{jl}(t,\tau)} dz 
\;
e^{i z^{2}}
\end{equation}

\be
w_{jl}(t,\tau) = \f{|y_{j}(t) - y_{l}(\tau)|}{2 \sqrt{t-\tau}}, \;\;
t> \tau
\end{equation}

\be
D_{jl}(t,\tau) = \f{\sqrt{- 2 i}}{\pi} \int_{\tau}^{t} d
\sigma \f{1}{\sqrt{t - \sigma}} U_{0}(\sigma - \tau ;y_{j}(\sigma) -
y_{l}(\tau))
\end{equation}

\n
A similar representation for the solution holds for $t<s$ (see Sect.
4).
\end{theorem}

\vspace{0.5cm}

\n
Using the representation of the solution  we can moreover prove

\vspace{0.5cm}

\begin{theorem}
\label{Th 2} The map $f \rightarrow \psi_{s}(t)$, $s,t \in R$,  
extends
uniquely to a unitary map $U(t,s)$ in $L^{2}(R^{3})$.
\end{theorem}

\vspace{0.5cm}

\n
The conditions we impose on the smoothness of the curves are not
optimal.
\n
Optimal conditions can be found analyzing in detail the representation
of the solution. We do not discuss further this problem here but 
notice that 
it may be relevant in the coupled case.

\vspace{1cm}

\section{Some auxiliary lemmas}

\vspace{0.5cm}

We shall  construct the solution of (12),(13) for $t\geq s$. The case
$t\leq s$ is obtained following the same steps  and it is outlined in
Sect. 4. We start
considering $\psi_{s}(t)$  given by  (14) for some functions 
$q_{j}(t)$.

\n
In the following we shall drop the dependence on the initial time $s$.

\n
We show first that if $q_{j}(t)$ and $y(t)$ are sufficiently smooth, e.g.
$y \in {\mathcal M}$ and $q_{j} \in W^{1,1}_{loc}(R)$, then $\psi(t)$
belongs
to the form domain $V_t$. It will also be clear that $\psi(t)$
does not belong to the operator domain even for an arbitrarly smooth
$q_{j}(t)$.

\n
In the second step, using the representation (14) for $\psi(t)$, we 
reduce
the solution of (12),(13) to an integro-differential equation for
$q_{j}(t)$.

\n
In the third step we show that the resulting equation is in fact 
equivalent
to the
integral equation (16), which has a unique solution with the required
regularity.

\n
The first result is summarized in the following lemma

\vspace{0.5cm}

\begin{lemma}
\label{Lemma1} Assume  $y \in {\mathcal M}$ and $q_{j} \in
W^{1,1}_{loc}(R)$,
with
$q_{j}(s)=0$ and $f \in C_{y(s)}^{\infty}$. Then
$\psi(t) \in V_t$, where $\psi(t)$ is given by (14).
\end{lemma}

\vspace{0.5cm}

{\em Proof.}  Expression (14) has a simpler form in the Fourier space

\be
\tilde{\psi} (k,t) = e^{- i k^{2} (t-s)} \tilde{f}(k) + \f{i}{(2
\pi)^{3/2}}
\sum_{j=1}^{n}\int_{s}^{t}
d\tau e^{- i k^{2} (t-\tau)} e^{- i k \cdot y_{j}(\tau)}
q_{j}(\tau)
\end{equation}

\n
We prove first that $\psi(t) \in L^{2}(R^{3})$. Due to the regularity
assumptions on $y$, $q_{j}$ and $f$,  it is sufficient to prove that

\be
\int_{|k|>1} dk |\zeta_{j}(k,t)|^{2} < \infty
\end{equation}

\n
where

\be
\zeta_{j}(k,t) =
 \f{i}{(2 \pi)^{3/2}} \int_{s}^{t} d\tau e^{- i k^{2} (t-\tau)}
e^{- i k \cdot y_{j}(\tau)} q_{j}(\tau)
\end{equation}

\n
An integration by parts yields

\ba
& &\zeta_{j}(k,t) =
- \f{1}{(2 \pi)^{3/2} k^{2}} \int_{s}^{t} d \tau e^{- i k^{2} (t-
\tau) - i k \cdot y_{j}(\tau)} \dot{q}_{j}(\tau) \nonumber\\
& & + \f{i}{(2 \pi)^{3/2} k^{2}} \int_{s}^{t} d \tau k \cdot
\dot{y}_{j}(\tau)  e^{- i k^{2} (t-
\tau) - i k \cdot y_{j}(\tau)} q_{j}(\tau) + \f{e^{- i k \cdot
y_{j}(t)} q_{j}(t)}{(2 \pi)^{3/2} k^{2}}
\ea

\n
The only delicate term in r.h.s. of (25) is the second. The explicit
computation of its $L^{2}$-norm gives

\ba
& &\f{1}{(2 \pi)^{3}} \int_{R^{3}} dk \f{1}{k^{4}} \int_{s}^{t} d
\tau  \int_{s}^{t} d \sigma q_{j}(\tau) \bar{q}_{j}(\sigma) k \cdot
\dot{y}_{j}(\tau)  k \cdot \dot{y}_{j}(\sigma) \nonumber\\
& &\times  e^{- i k^{2}
(\sigma - \tau) -i k \cdot (y_{j}(\tau) - y_{j}(\sigma)) } \nonumber\\
& & = \f{1}{(2 \pi)^{3}} \int_{s}^{t} d \tau \int_{s}^{t} d \sigma
q_{j}(\tau) \bar{q}_{j}(\sigma) |\dot{y}_{j}(\tau) |
|\dot{y}_{j}(\sigma) |\int_{S^{2}} d \Omega (\theta, \phi) \cos
\xi_{j}^{\tau}  \cos \xi_{j}^{\sigma} \nonumber\\
& & \times \int_{0}^{\infty} dk e^{- i
k^{2} ( \sigma - \tau) - i k |y_{j}(\tau) - y_{j}(\sigma) | \cos
\theta}
\ea

\n
where we have denoted by $\xi_{j}^{\nu}$ the angle between $k$ and
$\dot{y}_{j}(\nu)$. The last integral in (26)
can be written as

\ba
& &\int_{0}^{\infty} dk e^{- i
k^{2} ( \sigma - \tau) - i k |y_{j}(\tau) - y_{j}(\sigma) | \cos
\theta}\nonumber\\
& &= \f{e^{i \gamma^{2}}}{\sqrt{\sigma - \tau}}
\int_{\gamma}^{\infty} d z e^{- i z^{2}}, \;\; \gamma \equiv
\f{|y_{j}(\tau) - y_{j}(\sigma) | \cos \theta}{2
\sqrt{\sigma - \tau}}, \;\; \sigma > \tau
\ea

\n
 and similarly for $\tau > \sigma$.

\n
Using (27) one easily sees that the l.h.s. of (26) is finite and 
hence 
one 
concludes that $\psi(t) \in L^{2}(R^{3})$.

\n
Now we have to show that

\be
| \nabla \phi (t)| \equiv \left| \nabla \left(  \psi (t) - \sum_{j=1}^{n} 
q_{j}(t)
G(\cdot - y_{j}) \right) \right| \in L^{2} (R^{3})
\end{equation}

\n
From (22),(25) we have

\ba
& &\tilde{\phi}(k, t) = e^{- i k^{2} (t-s)} \tilde{f}(k)  - \f{1}{(2
\pi)^{3/2} k^{2}} \sum_{j=1}^{n} \int_{s}^{t} d \tau \dot{q}_{j}
(\tau) e^{- i k^{2} (t- \tau) - i k \cdot y_{j}(\tau)} \nonumber\\
& & + \f{i}{(2 \pi)^{3/2} k^{2}} \sum_{j=1}^{n} \int_{s}^{t} d \tau
q_{j}(\tau)
k \cdot \dot{y}_{j}(\tau)
 e^{- i k^{2} (t- \tau) - i k \cdot y_{j}(\tau)} \nonumber\\
& &\equiv \tilde{\phi}_{1}(k,t) + \tilde{\phi}_{2}(k,t) +
\tilde{\phi}_{3}(k,t)
\ea

\n
The smoothness of $f$ guarantees that $|\nabla \phi_{1}(t)| \in
L^{2}(R^{3})$. Concerning $\phi_{2}(t)$ we have

\ba
& &\int_{R^{3}} dk k^{2} | \tilde{\phi}_{2}(k,t) |^{2} \nonumber\\
& &\leq c \sup_{j} \int_{R^{3}} dk k^{2} \left| \f{1}{k^{2}}
\int_{s}^{t} d \tau \dot{q}_{j}(\tau)  e^{- i k^{2} (t- \tau) - i k 
\cdot
y_{j}(\tau)}
\right|^{2} \nonumber\\
& &= c \sup_{j} \int_{s}^{t} d \tau \int_{s}^{t} d \sigma
\dot{q}_{j}(\tau)  \overline{\dot{q}}_{j}(\sigma) \int_{R^{3}} dk
\f{1}{k^{2}}
e^{- i k^{2} (\sigma - \tau) + i k \cdot (y_{j}(\sigma) -
y_{j}(\tau) )}
\ea

\n
The last integral can be explicitely computed. Using spherical
coordinates and the position (20), for $\sigma > \tau$  one has

\ba
& &\int_{R^{3}} dk \f{1}{k^{2}} e^{- i k^{2} (\sigma - \tau) + i k 
\cdot
(y_{j}(\sigma ) - y_{l}(\tau) )} \nonumber\\
& &= \f{2 \pi}{\sqrt{\sigma - \tau} w_{jl}(\sigma , \tau)}
\int_{0}^{\infty} dp
\f{e^{-i p^{2}}}{p} \sin \left( 2 w_{jl}(\sigma , \tau)p \right)
\nonumber\\
& &=\f{2 \pi^{3/2}}{\sqrt{i} \sqrt{\sigma - \tau}} B_{jl}(\sigma 
,\tau)
\ea

\n
(See e.g. \cite{E}). An analogous computation holds for $\sigma < 
\tau$. 
The 
function $B_{jl}(t,s)$, $t>s$, has been defined in (19) and it is
continuous in both variables and differentiable in the second one.

\n
From (30) and (31) one easily gets the estimate for $\phi_{2}$.

\n
It remains to estimate $\phi_{3}(t)$. A further integration by parts
yields

\ba
& &(2 \pi)^{3/2} \tilde{\phi}_{3} (k,t) =
 \f{i}{k^{4}} \sum_{j=1}^{n} \int_{s}^{t} d \tau
\dot{q}_{j}(\tau)  k \cdot \dot{y}_{j}( \tau)
e^{-i k^{2} (t- \tau) - i k \cdot y_{j}(\tau) } \nonumber\\
& &+ \f{i}{k^{4}} \sum_{j=1}^{n} \int_{s}^{t} d \tau
q_{j}(\tau) k \cdot \ddot{y}_{j}(\tau)
e^{-i k^{2} (t- \tau) - i k \cdot y_{j}(\tau) }\nonumber\\
& &+ \f{1}{k^{4}} \sum_{j=1}^{n} \int_{s}^{t} d \tau
q_{j}(\tau) \left( k \cdot \dot{y}_{j} (\tau) \right)^{2}
e^{-i k^{2} (t- \tau) - i k \cdot y_{j}(\tau) }\nonumber\\
& &- \f{i}{k^{4}} \sum_{j=1}^{n} q_{j}(t) k \cdot \dot{y}_{j}(t)
e^{-i k \cdot y_{j}(t)}
\ea

\n
The only delicate term in the r.h.s. of (32) is the third one. 
Proceeding
as
in (30), one easily sees that its gradient has a finite $L^{2}$-norm
and this concludes the proof of the lemma.

\vspace{0.5cm}
\n
{\bf Remark}. From (29), (32) we get the following representation
for $\tilde{\phi} (k,t)$

\[ \tilde{\phi} (k,t) = \tilde{\chi} (k,t) - \f{i}{(2 \pi)^{3/2}
k^{4}} \sum_{j=1}^{n} q_{j}(t) k \cdot \dot{q}_{j} (t) e^{-i k \cdot
y_{j}(t)}  \]

\n
Assuming further regularity on $y$, $q_{j}$ and using again
integration by parts one easily sees that $\Delta \chi \in
L^{2}(R^{3})$ which implies $\Delta \phi \not\in L^{2}(R^{3})$.

\n
This means that $\psi(t)$ given by (22) does not belong to
$D(H_{y(t)})$, i.e. problem (1) does not have strong solutions.

\vspace{0.5cm}

\n
In the next lemma we reduce the evolution problem to the solution of 
an
integro-differential equation for $q_{j}(t)$.

\vspace{0.5cm}

\begin{lemma}
\label{Lemma1} Assume $y \in {\mathcal M}$, $q_{j} \in W^{1,1}_{loc}
(R)$, with $q_{j}(s) = 0$, and $f \in C_{y(s)}^{\infty}$. Then 
$\psi(t)$
given by (14) solves problem  (12) (13) if $q_{j}(t)$ solves the
equation
\end{lemma}

\ba
& & 4\pi (U_{0}(t-s)f)(y_{j}(t))=
4 \pi\alpha_{j} q_{j}(t) - \sum_{l=1,l\neq j}^{n}
\f{q_{l}(t)}{|y_{j}(t)- y_{l}(t)|} \nonumber\\
& & + \f{1}{\sqrt{i\pi}} \sum_{l=1}^{n} \int_{s}^{t} d\tau
\dot{q}_{l}(\tau)
\f{B_{jl}(t,\tau)}{\sqrt{t-\tau}} - \f{\sqrt{i}}{\sqrt{\pi}}
\sum_{l=1}^{n}\int_{s}^{t} d\tau q_{l}(\tau)
\f{A_{jl}(t,\tau)}{\sqrt{t-\tau}}
\nonumber\\
& &
\ea

\n
{\em where $B_{jl}(t,\tau)$ and $A_{jl}(t,\tau)$ are given in
(19),(18).}

\vspace{0.5cm}

{\em Proof.} From lemma 3.1 we know that $\psi(t) \in V_t$ and then 
the r.h.s.
of (12) is well defined.  Now we check  that $\f{\partial \psi (t
)}{\partial t} \in V^{*}_{t}$.
  A direct computation yields (see (22),(25))

\ba
& &\f{\partial \tilde{\psi}(k,t)}{\partial t} = - i k^{2} e^{-i 
k^{2}(t-s)}
\tilde{f}(k) + i \sum_{j=1}^{n}  \int_{s}^{t} d\tau
e^{- i k^{2} (t-\tau)} \f{d}{d\tau} \left( q_{j}(\tau) \f{e^{- i k 
\cdot
y_{j}(\tau)}}{(2
\pi)^{3/2}}
\right)
\nonumber\\
& &= - i k^{2} \left[ e^{- i k^{2} (t-s)} \tilde{f} (k) - \f{1}{(2
\pi)^{3/2} k^{2}} \sum_{j=1}^{n} \int_{s}^{t} d \tau e^{-i k^{2}
(t-s)} \f{d}{d \tau} \left( q_{j}(\tau) e^{-i k \cdot y_{j}(\tau)}
\right) \right] \nonumber\\
& &= -i k^{2} \tilde{\phi}(k,t)
\nonumber\\
& &
\ea

\n
For $v(t) \in V_t$, we write

\be
v(t) = \xi^{v} (t) + \sum_{j=1}^{n} q^{v}_{j} (t) G( \cdot - 
y_{j}(t)) \;
, \;\;\;\;
\xi^{v} (t) \in H^{1}(R^{3})
\end{equation}

\n
Using (34) we have

\be
\left( \xi^{v} (t) , \f{\partial \psi(t)}{\partial t} \right) = - i
\int_{R^{3}} dx \nabla \bar{\xi^{v}}(x,t) \cdot \nabla \phi (x,t)
\end{equation}

\n
which is obviously finite. Moreover

\ba
& &\left( G(\cdot - y_{j}(t)) , \f{\partial \psi(t)}{\partial t}
\right) =
\int_{R^{3}} dk \f{e^{i k \cdot y(t)}}{(2 \pi)^{3/2}k^{2}} \f{\partial
\tilde{\psi}(k,t)}{\partial t}\nonumber\\
& &=\int_{R^{3}} dk \f{e^{i k \cdot y_{j}(t)}}{(2 \pi)^{3/2}k^{2}}
\nonumber\\
& &\times \left[ - i
k^{2}
e^{-i k^{2} (t-s)} \tilde{f}(k) +
\f{i}{( 2 \pi)^{3/2}}  \sum_{l=1}^{n}  \int_{s}^{t} d \tau e^{-i k^{2}
(t-\tau)}
 \f{d}{d\tau} \left( q_{l}(\tau)
e^{-i k \cdot y_{l}(\tau)} \right) \right] \nonumber\\
& &= - i (U_{0}(t-s) f)(y_{j}(t)) \nonumber\\
& &+ \f{i}{(2 \pi)^{3}}
\sum_{l=1}^{n} \int_{s}^{t} d \tau \dot{q}_{l}(\tau)
\int_{R^{3}} dk \f{1}{k^{2}} e^{- i k^{2} (t- \tau) + i k \cdot
(y_{j}(t) - y_{l}(\tau) )} \nonumber\\
& & +\f{1}{(2 \pi)^{3}} \sum_{l=1}^{n} \int_{s}^{t} d \tau q_{l}(
\tau) \int_{R^{3}} dk \f{k \cdot \dot{y}_{l}(\tau)}{k^{2}}
 e^{- i k^{2} (t- \tau) + i k \cdot
(y_{j}(t) - y_{l}(\tau) )}
\ea

\n
The last integral in the $k$-variable can be explicitely computed. We
introduce spherical coordinates $k=(r, \theta, \phi)$ with polar  
axis directed along 
$y_{j}(t) - y_{l}(\tau)$ and 
$\dot{y}_{l}(\tau) = (|\dot{y}_{l}(\tau)|, \hat{\theta}, 0)$. Using 
the
formula

\be
k \cdot \dot{y}_{l}(\tau) = r |\dot{y}_{l}(\tau)| \left( \cos \theta 
\cos
\hat{\theta}+ \sin \theta \sin \hat{\theta} \cos \phi \right)
\end{equation}

\n
we have

\ba
& &\int_{R^{3}} dk \f{k \cdot \dot{y}_{l}(\tau)}{k^{2}}
 e^{- i k^{2} (t- \tau) + i k \cdot
(y_{j}(t) - y_{l}(\tau) )}\nonumber\\
& &=\f{4 \pi |\dot{y}_{l}(\tau)| \cos \hat{\theta}}{i \sqrt{t- \tau}
|y_{j}(t) - y_{l}(\tau) |} \left[ \int_{0}^{\infty} dp e^{-i p^{2}}
\cos (2 w_{jl}(t,\tau) p)\right. \nonumber\\
& &\left.  - \f{1}{2 w_{jl}(t, \tau)}
\int_{0}^{\infty} dp \f{e^{-i p^{2}}}{p} \sin (2 w_{jl}(t, \tau) p)
\right]\nonumber\\
& &= \f{2 (\pi)^{3/2} |\dot{y}_{l}(\tau)| \cos \hat{\theta}}{i 
\sqrt{i}
 \sqrt{t- \tau}
|y_{j}(t) - y_{l}(\tau) |}
\left[ e^{i w_{jl}(t,\tau)^{2}} - \f{1}{w_{jl}(t,\tau)}
\int_{0}^{w_{jl}(t,\tau)} dv e^{i v^{2}} \right] \nonumber\\
& &\f{4 (\pi)^{3/2} |\dot{y}_{l}(\tau)| \cos \hat{\theta}}{\sqrt{i}
 \sqrt{t- \tau} |y_{j}(t) - y_{l}(\tau) |} \f{1}{w_{jl}(t,\tau)}
\int_{0}^{w_{jl}(t,\tau)} dv v^{2} e^{i v^{2}} \nonumber\\
& &=\f{2 \pi^{3/2}}{\sqrt{i} \sqrt{t-\tau} } A_{jl}(t,\tau)
\ea

\n
where the function $A_{jl}(t, \tau)$ has been defined in (18).
From (37), (39)  we find

\ba
& &\left( G(\cdot - y_{j}(t)) , \f{\partial \psi(t)}{\partial t}
\right) = - i (U_{0}(t-s)f)(y_{j}(t) \nonumber\\
& &+ \f{\sqrt{i}}{4 \pi^{3/2}} \sum_{l=1}^{n} \int_{s}^{t} d \tau
\dot{q}_{l}(\tau) \f{B_{jl}(t,\tau)}{\sqrt{t-\tau}}
+\f{1}{4 \sqrt{i} \pi^{3/2}} \sum_{l=1}^{n} \int_{s}^{t} d \tau
q_{l}(\tau) \f{A_{jl}(t,\tau)}{\sqrt{t-\tau}}\nonumber\\
& &
\ea

\n
which implies $\f{\partial \psi (t)}{\partial t} \in V^{*}_{t}$.
Using (36),(40), it is now easy to check that the evolution equation
(12)  reduces to
the integro-differential equation (33) which is satisfied by 
hypotheses.
This concludes the proof of  lemma 3.2.

\vspace{1cm}

\section{Proof of theorem 2.1}

\vspace{0.5cm}

\n
We shall use the results of lemma 3.1 and 3.2   to complete the 
proof of theorem 2.1.

\n
Let us  fix $y \in {\mathcal M}$ and  the initial datum $f \in
C_{y(s)}^{\infty}$. For $t>s$  we consider the
equation
(16) for
$q_{j}(t)$.

\n
It is a Volterra integral equation containing the Abel operator

\be
\left( L q_{j} \right)(t) \equiv \f{1}{\sqrt{-i\pi}}  \int_{s}^{t}
d\tau \f{q_{j}(\tau)}{\sqrt{t-\tau}}
\end{equation}

\n
and the integral operators

\be
\left( C_{j} q_{j} \right)(t) \equiv \int_{s}^{t} d \tau
q_{j}(\tau) C_{j}(t,
\tau)
\end{equation}

\be
\left( D_{jl} q_{l} \right) \equiv \int_{s}^{t} d \tau
q_{l}(\tau) D_{jl}(t, \tau) \;\;\;\;\;  j \neq l
\end{equation}

\n
The datum of the equation

\be
h_{j} (t) \equiv \f{4 \sqrt{\pi}}{\sqrt{-i}}
\int_{s}^{t} d\tau \f{(U_{0}(\tau - s) f)(y_{j}(\tau))}{\sqrt{t-\tau}}
\end{equation}

\n
is the result of the application of the Abel operator $L$ to the
function

\n
$4 \pi (U_{0}(t - s) f)(y_{j}(t)) $. Due to the smoothness of $f$ it 
is obviously true that $h_{j} \in
W^{1,1}_{loc}(R)$ and $h_{j}(s)=0$.

\n
By direct inspection of (17)-(20), one verifies that if $y \in
{\mathcal M}$  then $C_{j}(t,\tau)$ is continuous in both variables 
and it
is  differentiable
 as a function of $t$.

\n
By a detailed analysis of  the expression (21) (which we omit for
brevity), one can also show that $D_{jl}$ is a bounded operator in
$W^{1,1}_{loc}(R)$.

\n
The same is true for the Abel operator (see e.g. \cite{GV}) and 
we  conclude that
equation (16) has a unique solution $q_{j} \in W^{1,1}_{loc}(R)$, with
$q_{j}(s)=0$.

\n
Now we apply the Abel operator to equation (16), and make use of
the fact
 that for $\eta $ differentiable with $ \eta (s) = 0$
one has

\be
\f{d}{dt}[(L)^{2}\eta] = i \eta, \;\;  \;\;
\f{d}{dt}(L \eta )(t) = (L \dot{\eta} )(t)
\end{equation}

\n
The resulting equation reads

\ba
& &- i (L\dot{q}_{j})(t) + 4\pi \alpha_{j}  q_{j}(t)  - i
\f{d}{dt}(L C_{j} q)(t)  - i  \sum_{l=1, l\neq j}^{n}
\f{d}{dt} \left( L D_{jl} q_{l}\right) (t)\nonumber\\
& &= 4 \pi  (U_{0}(t - s) f)(y_{j}(t))
\ea

\n
The integral operator $C_{j}$ can be rewritten as

\ba
& &(C_{j} q_{j})(t) \nonumber\\
& &= - \f{1}{\pi}\int_{s}^{t} d\sigma
\f{1}{\sqrt{t-\sigma}}\int_{s}^{\sigma}d\tau q_{j}(\tau) \left(
\f{i A_{jj}(\sigma, \tau)}{\sqrt{\sigma-\tau}} + \f{d}{d\tau}\left(
\f{B_{jj}(\sigma,\tau) - 1}{
\sqrt{\sigma  - \tau}}\right) \right) \nonumber\\
& &= - \f{i}{\pi}\int_{s}^{t} d\sigma
\f{1}{\sqrt{t-\sigma}}\int_{s}^{\sigma}d\tau q_{j}(\tau)
\f{A_{jj}(\sigma, \tau)}{\sqrt{\sigma-\tau}}\nonumber\\
& & + \f{1}{\pi}\int_{s}^{t} d\sigma
\f{1}{\sqrt{t-\sigma}}\int_{s}^{\sigma}d\tau \dot{q}_{j}(\tau)
\f{B_{jj}(\sigma,\tau) - 1}{\sqrt{\sigma  - \tau}}
\ea

\n
Using again the first equation  in (45)  we
have

\ba
& &\f{d}{dt}(L C_{j}q_{j})(t) =  \f{\sqrt{-i}}{\sqrt{\pi}}\int_{s}^{t}
d\tau
q_{j}(\tau)
\f{A_{jj}(t, \tau)}{\sqrt{t-\tau}} + i \f{\sqrt{-i}}{\sqrt{\pi}}
\int_{s}^{t}d\tau \dot{q}_{j}(\tau)
\f{B_{jj}(t,\tau)}{\sqrt{t - \tau}}\nonumber\\
& &- i \f{\sqrt{- i }}{\sqrt{\pi}} \int_{s}^{t} d \tau
\f{\dot{q}_{j}(\tau)}{\sqrt{t- \tau}}
\ea

\n
Concerning the integral operator $D_{jl}$ we have

\ba
& & \left( L D_{jl} q_{l} \right) (t) = \sqrt{\f{2}{\pi}}
 \int_{s}^{t} d
\tau q_{l}(\tau) \int_{\tau}^{t} d \sigma U_{0}(\sigma - \tau ;
y_{j}(\sigma) - y_{l}(\tau)) \nonumber\\
& &= \f{1}{2 \pi^{2}} \int_{s}^{t} d \sigma \int_{R^{3}} dk
\int_{s}^{\sigma} d \tau
q_{l}(\tau) e^{- i k^{2} (\sigma - \tau) + i (y_{j}(\sigma) -
y_{l}(\tau))} \nonumber\\
& & =\f{i}{2 \pi^{2}} \int_{s}^{t} d \sigma \int_{s}^{\sigma} d \tau
\int_{R^{3}}d k
\f{1}{k^{2}} \f{d}{d \tau} \left( q_{l}(\tau) e^{- i k \cdot
y_{l}(\tau)} \right) e^{- i k^{2} (\sigma - \tau) + i k \cdot
y_{j}(\sigma) }\nonumber\\
& &- \f{i}{2 \pi^{2}} \int_{s}^{t} d \sigma q_{l}(\sigma) 
\int_{R^{3}}dk
\f{1}{k^{2}} e^{i k \cdot (y_{j}(\sigma) - y_{l}(\sigma))} \nonumber\\
& &=\f{i}{\sqrt{i} \sqrt{\pi}}  \int_{s}^{t} d \sigma 
\int_{s}^{\sigma} d
\tau \dot{q}_{l}(\tau)
\frac{B_{jl}(\sigma, \tau)}{\sqrt{\sigma - \tau}} +
\f{1}{\sqrt{i} \sqrt{\pi}} \int_{s}^{t} d \sigma \int_{s}^{\sigma} d 
\tau
q_{l}(\tau)
\f{A_{jl}(\sigma , \tau)}{\sqrt{\sigma - \tau}} \nonumber\\
& &- i \int_{s}^{t} d \sigma q_{l}(\sigma) \f{1}{|y_{l}(\sigma) -
y_{j}(\sigma)|}
\ea

\n
If we substitute (49),(48) into equation (46) we find that the charges
$q_{j}(t)$ satisfy  the integro-differential equation (33). By lemma 
3.2
this means that if the $q_{j}(t)$ solve equation (16) then $\psi(t)$
given in (14) solves the evolution problem (12),(13) for $t>s$.

\n
We now briefly consider the case of the backward evolution. More
precisely, given the initial time $t\in R$ and $g \in
C^{\infty}_{y(t)}$, we want to find $\psi_{t}(s) \in V_{s}$, for
$s<t$, satisfying the equation

\ba
& &i \left( v(s) , \f{\partial \psi_{t}(s)}{\partial s} \right) =
{\mathcal B}_{y(s)} (v(s), \psi_{t}(s)) \;\;\;\; \forall v(s) \in
V_{s}\nonumber\\
& &\psi_{t}(t)=g
\ea

\n
Again we start representing $\psi_{t}(s)$ as

\be
\psi_{t}(s) = U_{0}^{*}(t-s) g - i \sum_{j=1}^{n} \int_{s}^{t} d \tau
U_{0}^{*} (\tau - s ; \cdot - y_{j}(\tau)) \tilde{q}_{j}(\tau)
\end{equation}

\n
for some functions $\tilde{q}_{j}$ and then we determine
$\tilde{q}_{j}$ in such a way that (51) solves (50).

\n
The steps are  similar to the case of the forward evolution and will 
be
omitted. We only write the integral equation which is satisfied by
$\tilde{q}_{j}$

\ba
& &\tilde{q}_{j}(s) + \alpha_{j} \f{4 \sqrt{\pi}}{\sqrt{i}}
\int_{s}^{t} d \tau \f{\tilde{q}_{j}(\tau) }{\sqrt{\tau -s}} +
\int_{s}^{t} d \tau \tilde{q}_{j}(\tau) \bar{C}_{j}( \tau,
s)\nonumber\\
& &+ \sum_{l=1,l \neq j}^{n} \int_{s}^{t} d \tau \tilde{q}_{l}(\tau)
\bar{D}_{jl}(\tau, s) = \f{4 \sqrt{\pi}}{\sqrt{i}} \int_{s}^{t} d \tau
\f{(U_{0}^{*} (t - \tau) g)(y_{j}(\tau) )}{\sqrt{\tau -s}}\nonumber\\
& &
\ea

\n
Finally  the  uniqueness of the solution of problem (12), (13)  easily
follows
 from the fact that  for any solution of (12) the
$L^{2}$-norm  is conserved.

\vspace{1cm}

\section{Unitary evolution}

\vspace{0.5cm}

In this section we give the proof of theorem 2.2 following the idea
developped in \cite{SY} for the case of point interactions at fixed
positions with time-dependent strengths.

\n
We fix $s,t \in R$ and, without loss of generality, we take $s \leq 
t$.

\n
By theorem 2.1 we have existence and uniqueness of the forward
evolution $\psi_{s}(t)$ and the backward evolution $\psi_{t}(s)$ for
smooth initial data, denoted respectively  by $f$ and $g$.

\n
Moreover the linear maps

\be
f \rightarrow \psi_{s}(t),  \;\;\;\;  g \rightarrow \psi_{t}(s)
\end{equation}

\n
are both defined on a dense set of $L^{2}(R^{3})$ and are isometries.
Then they can be uniquely extended to isometries on $L^{2}(R^{3})$.

\n
We shall denote them respectively by $U(t,s)$ and $U(s,t)$.

\n
In order to prove that they are unitary maps, we have to show
that the adjoints $U^{*}(t,s)$  and $U^{*}(s,t)$ are also
isometries. This fact will follow from the equalities

\be
U^{*}(t,s) = U(s,t), \;\;\;\; U^{*}(s,t)= U(t,s)
\end{equation}

\n
Here we prove the first equality in (54) (the second is obtained in
the same way).

\n
For the sake of clarity, it is convenient to rewrite $U(t,s)$ using
various integral operators. In particular we use the Abel operator $L$
on $L^{2}([s,t])$ defined in (41) and its adjoint $L^{*}$, the
operator $T \;: \; L^{2}(R^{3}) \rightarrow \oplus_{j=i}^{n} 
L^{2}([s,t])$  defined by

\be
(Th)_{j}(\tau)  = 2 \sqrt{\pi} \int_{R^{3}} dx U_{0}(\tau ; x -
y_{j}(\tau)) h(x)
\end{equation}

\n
and its adjoint $T^{*}$. Moreover, from (17),(21), we have

\be
C_{j}=LR_{j}, \;\;\;\;\; D_{jl}=LS_{jl}
\end{equation}

\n
where $R_{j}$ and $S_{jl}$ are the integral operators defined by

\ba
& &(R_{j}q_{j})(\tau) = -\f{1}{\sqrt{i \pi}} \int_{s}^{\tau} d \sigma
\f{q_{j} (\sigma)}{\sqrt{\tau - \sigma}} \left( i A_{jj}(\tau ,
\sigma) + \f{dB_{jj}}{d \tau}(\tau, \sigma) + \f{B_{jj}(\tau,\sigma)
-1}{2(\tau - \sigma)} \right) \nonumber\\
& &
\ea

\be
(S_{jl} q_{j})(\tau) = - \f{i}{\sqrt{\pi}} \int_{s}^{\tau} d \sigma 
q_{j}
(\sigma) U_{0}(\tau - \sigma ; y_{j}(\tau) - y_{l}(\sigma))
\end{equation}

\n
We also introduce   the corresponding operators on the
space of vector valued functions ${\bf q}(t) = (q_{1}(t),
\ldots , q_{n}(t))$, i.e.

\ba
& &(R {\bf q})_{j}(\tau) = (R_{j}q_{j})(\tau)\\
& &(S {\bf q})_{j}(\tau) = \sum_{l=1, l \neq j}^{n}
(S_{jl}q_{l})(\tau) \\
& &(\Lambda {\bf q})_{j}(\tau) = 4 \pi \alpha_{j} q_{j}(\tau)
\ea

\n
Using the above notation we rewrite equation (16) in the form

\be
{\bf q} + L( \Lambda + R + S) {\bf q} = 2 \sqrt{\pi} L T U_{0}^{*}(s) 
f
\end{equation}

\n
From (14), (55) and (62) we obtain the following representation of
$U(t,s)f$,  for $s \leq t$ and $f \in C^{\infty}_{y(s)}$

\be
U(t,s)f = U_{0}(t-s)f + i U_{0}(t) T^{*} \left[ I + L (\Lambda + R +
S) \right]^{-1} L T U_{0}^{*} (s) f
\end{equation}

\n
The same procedure can also be applied to obtain the representation
of $U(s,t)$, for $s \leq t$ and $g \in C^{\infty}_{y(t)}$

\be
U(s,t) g= U_{0}^{*}(t-s) g - i U_{0}(s) T^{*} \left[ I + L^{*} ( 
\Lambda +
R^{*} + S^{*} ) \right] ^{-1} L^{*} T U_{0}^{*}(t) g
\end{equation}

\n
A straightforward computation gives

\ba
& &(U^{*}(t,s)g, f)= (g, U(t,s)f)\nonumber\\
& &=( U_{0}^{*}(t-s) g,f) + i \left( U_{0}(s) T^{*} L^{*} \left[ (I +
L(\Lambda + R + S) )^{-1} \right]^{*} T U_{0}^{*} (t) g , f\right)
\nonumber\\
& &=( U_{0}^{*}(t-s) g,f) + i \left( U_{0}(s) T^{*} \left[ I + L^{*}
(\Lambda +R^{*} + S^{*} ) \right]^{-1} L^{*} T U_{0}^{*}(t)g, f\right)
\nonumber\\
& &
\ea

\n
Thus, from (64), (65),  we have $U^{*}(t,s) = U(s,t)$ on a dense set 
and
then on
$L^{2}(R^{3})$  and this concludes the proof of the theorem.

\vspace{1cm}
\n \textbf{Acknowledgements.} One of us (G.F.D.A.) is grateful to the
Von Humboldt Foundation for the generous support.


\vspace{1cm}
\begin{thebibliography}{99}

\bibitem{AGH-KH} Albeverio S., Gesztesy F., Hogh-Krohn R. and Holden 
H.,
Solvable Models in Quantum Mechanics, Springe-Verlag, New-York, 1988.

\bibitem{DFT1} Dell'Antonio G.F., Figari R. and Teta A., A limit 
evolution
problem for time-dependent point interactions, {\em J. Func.
Anal.}, {\bf 142}, 1996, 249-275.


\bibitem{DFT2} Dell'Antonio G.F., Figari R. and Teta A., Diffusion of 
a
particle in presence of N moving point sources, {\em Ann. Inst. H.
Poincar\'{e} A}, {\bf 69}, 1998, 413-424.

\bibitem{E} Erdely et al., Tables of Integral Transforms, vol. 1,
McGraw-Hill,
New-York, 1954.

\bibitem{GV} Gorenflo R. and Vessella S., Abel Integral Equation,
Springer-Verlag, Berlin-Heidelberg, 1991.

\bibitem{SY} Sayapova M.R. and Yafaev D.R., The evolution operator for
time-dependent potentials of zero radius, {\em Proc. Stek. Inst.
Math.}, {\bf 2}, 1984, 173-180.

\bibitem{T} Teta A, Quadratic forms for singular perturbation of the
laplacian, {\em Publ. R.I.M.S.}, {\bf 26}, 1990, 803-817.

\bibitem{Y} Yafaev D.R., Scattering theory for time-dependent 
zero-range
potentials, {\em Ann. Inst. H. Poincare' A}, {\bf 40}, 1984, 343-359.

\bibitem{Ya} Yajima K., Existence of solutions for Schr\"{o}dinger 
evolution

equations, {\em Comm. Math. Phys.}, {\bf 110}, 1987, 415-426.
\end{thebibliography}
\end{document}